%% file: main.tex
\documentclass{article}

\usepackage[preprint]{neurips_2026}

\usepackage[utf8]{inputenc} 
\usepackage[T1]{fontenc}    
\usepackage{hyperref}       
\usepackage{url}            
\usepackage{booktabs}       
\usepackage{amsfonts}       
\usepackage{nicefrac}       
\usepackage{microtype}      
\usepackage{xcolor}         
\usepackage{kotex}
\usepackage{amsmath}
\usepackage{amssymb}
\usepackage{graphicx}
\usepackage{subcaption}
\usepackage{multirow}

\title{TLDR: Compressing Audio Tokens for Efficient Autoregressive Text-to-Speech}  

\author{%
  Yejin Lee\textsuperscript{1}, Junwon Moon\textsuperscript{1}, Hyoeun Kim\textsuperscript{1}, Hyunjin Choi\textsuperscript{1}, Heeseung Kim\textsuperscript{2}, Kyuhong Shim\textsuperscript{1} \\
  \textsuperscript{1} Sungkyunkwan University, \textsuperscript{2} University of Seoul\\
  \texttt{\{yj.lee, mppn98, khshim\}@skku.edu}, \texttt{gmltmd789@uos.ac.kr} \\
}

\begin{document}

\maketitle

\begin{abstract}
\input{sections/00_abstract}
\end{abstract}

\input{sections/01_introduction}
\input{sections/02_related}
\input{sections/03_method}
\input{sections/04_experiment}
\input{sections/05_analysis}

\input{sections/06_conclusion} 


{
    \small
    \bibliographystyle{plain}
    \bibliography{reference}
}


\newpage
\appendix
\input{sections/Y_appendix}



\end{document}

%% file: sections/00_abstract.tex
Codec-based autoregressive (AR) speech language models have achieved strong text-to-speech (TTS) quality by modeling speech as sequences of discrete audio tokens with large pretrained backbones.
However, this token-level formulation creates a structural efficiency bottleneck: speech-token sequences are much longer than text sequences, requiring the AR backbone to perform causal computation at every token position and maintain a KV cache that grows with the sequence length.
We introduce TLDR, a patch-based autoregressive framework that accelerates codec-based AR-TTS by shifting the causal modeling from token-level speech sequences to patch-level sequences.
TLDR groups consecutive codec tokens into compact latent patches using a lightweight compressor, models the resulting shorter patch sequence with a frozen pretrained AR-TTS backbone adapted by LoRA, and reconstructs fine-grained speech tokens within each patch using a speaker-conditioned extractor.
With a patch size of 4, TLDR achieves a 1.8x inference speedup over the baseline AR-TTS model and reduces global KV-cache memory by up to 75\%.
Experimental results indicate that patch-level global causal modeling can be a practical way to reduce the inference cost of pretrained codec-based AR-TTS systems without replacing the existing modules.

%% file: sections/01_introduction.tex
\section{Introduction}\label{sec:intro}

\begin{figure}[!t]
\centering
\includegraphics[width=1\textwidth]{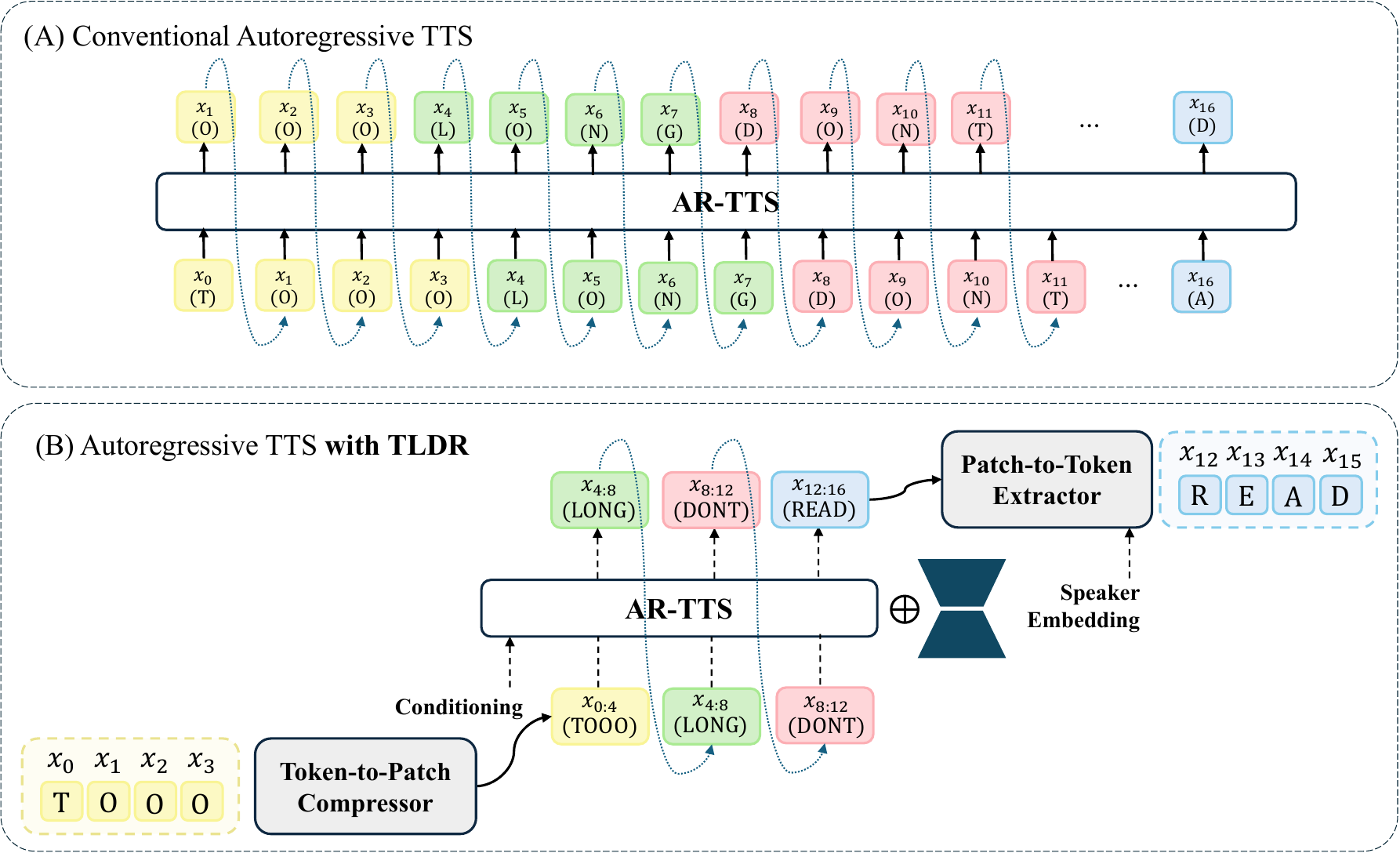} 
\caption{Overview of \textsc{TLDR}, a patch-based autoregressive framework for speech generation. 
A token-to-patch compressor maps consecutive codec tokens into patch representations, a patch-level global sequence model predicts the patch sequence, and a patch-to-token extractor generates fine-grained codec tokens within each patch.}
\label{fig:overview}
\vspace{0.3cm}
\end{figure}   

Codec-based autoregressive text-to-speech (AR-TTS) systems generate speech as sequences of discrete audio tokens produced by neural audio codecs~\cite{defossez2022high, zeghidour2021soundstream}.
This discretization recasts speech generation as next-token prediction over a finite vocabulary, replacing direct waveform modeling with discrete symbolic prediction~\cite{chen2024vall, guo2024fireredtts, wang2023neural, ye2025llasa, zhou2025indextts2}.
Consequently, AR-TTS systems predict symbolic acoustic units with a standard categorical objective instead of directly estimating high-dimensional waveform samples, yielding a tractable formulation for speech generation.

Despite this advantage, this token-level formulation imposes an inference bottleneck.
Because codec tokenizers encode speech at a high frame rate, even a short utterance can comprise hundreds or thousands of codec tokens.
Conventional AR-TTS systems therefore require one forward pass through the autoregressive decoder for each generated codec token.
Thus, the main efficiency challenge arises not only from long codec-token sequences, but also from requiring the autoregressive decoder to predict at every codec-token timestep.

Existing acceleration methods, such as multi-token prediction (MTP)~\cite{gloeckle2024better,nguyen2025accelerating,wang2025vocalnet} and speculative decoding (SD)~\cite{chen2023accelerating, leviathan2023fast}, accelerate token-level autoregressive decoding by predicting multiple future tokens or verifying draft predictions more efficiently.
However, they retain codec-token-level modeling: the AR-TTS model still predicts at codec-token timesteps, and its KV cache remains tied to the length of the generated codec-token sequence. 

In this paper, we propose \textbf{TLDR}, a patch-based autoregressive TTS framework that shifts the modeling unit of a pretrained AR-TTS backbone from individual codec tokens to codec-token patches.
TLDR partitions consecutive codec tokens into patches and feeds the resulting patch sequence to the pretrained AR-TTS backbone, which serves as the global sequence model.
A lightweight token-to-patch compressor maps codec-token spans into patch representations, while a speaker-conditioned patch-to-token extractor generates fine-grained codec tokens within each patch.

The key assumption behind TLDR is that codec-token sequences are locally redundant but globally structured.
Local redundancy arises because adjacent codec frames describe temporally contiguous speech segments whose acoustic attributes typically vary smoothly over short time spans.
Global structure, in contrast, is governed by slower dependencies such as text--speech alignment, prosodic progression, and speaker-consistent generation.
This distinction motivates reducing the update rate of the largest causal model from codec-token timesteps to patch timesteps.
Accordingly, TLDR compresses local token spans into patch representations for global autoregressive modeling and reconstructs token-level detail within each patch using a lightweight speaker-conditioned extractor.

By requiring one global-backbone forward pass per patch rather than per codec token, TLDR reduces the number of autoregressive decoding steps performed by the global sequence model.
The resulting shorter patch sequence also reduces the KV cache maintained by the global backbone.
Although TLDR still performs autoregressive decoding within each patch, this local decoding is bounded by the patch size $k$ and is handled by the patch-to-token extractor rather than the pretrained backbone.
The main efficiency gain therefore comes from reducing the update frequency and KV-cache growth of the global causal model from the token rate to the patch rate.

Our contributions are summarized as follows:
\begin{itemize}
\item We recast codec-based AR-TTS as patch-level autoregressive modeling, reducing the update frequency of the global causal backbone from codec-token timesteps to patch timesteps.

\item We retrofit a pretrained AR-TTS backbone to this patch-level formulation by freezing its base weights and training LoRA adapters together with a token-to-patch compressor and a speaker-conditioned patch-to-token extractor, while retaining the original tokenizer, text frontend, and vocoder.
 
\item We quantify the quality--latency--memory trade-off induced by patch size in zero-shot TTS, showing that TLDR reduces global-backbone KV-cache memory by approximately 75\% and achieves a 1.8$\times$ speedup at $k=4$ with minimal degradation in recognition accuracy and speaker similarity.

\end{itemize}

%% file: sections/02_related.tex
\section{Related Work}\label{sec:related}

\subsection{Autoregressive and Non-Autoregressive Text-to-Speech} 
Codec-token-based AR-TTS models represent speech as discrete audio-token sequences produced by neural audio codecs and formulate speech synthesis as conditional next-token prediction.
This formulation underlies recent zero-shot TTS systems built on neural audio codecs such as SoundStream~\cite{zeghidour2021soundstream} and EnCodec~\cite{defossez2022high}, as well as audio language modeling frameworks such as AudioLM~\cite{borsos2023audiolm}.
VALL-E~\cite{wang2023neural} instantiated this approach by autoregressively generating codec tokens from text and a short speech prompt.
Recent systems, including the CosyVoice series~\cite{du2024cosyvoice, du2024cosyvoice2, du2025cosyvoice3}, Seed-TTS~\cite{anastassiou2024seed}, and LLaSA~\cite{ye2025llasa}, further demonstrate the effectiveness of codec-token autoregression for zero-shot TTS.
Despite these advances, codec-token-based AR-TTS systems remain latency-intensive because each generated codec token requires a sequential autoregressive decoding step.

In parallel, non-autoregressive (NAR) TTS models~\cite{ren2020fastspeech2, ren2019fastspeech}, including diffusion- and flow-matching-based architectures~\cite{chen2025f5, eskimez2024e2, kim2020glowtts, le2023voicebox, mehta2024matcha, wang2024maskgct}, reduce latency by removing or shortening autoregressive dependency chains.
However, many NAR systems rely on duration or alignment estimation, masking schedules, or iterative denoising/flow procedures, which differ from causal codec-token autoregression.
TLDR instead targets efficiency within the AR-TTS paradigm: it preserves autoregressive generation but shifts the global modeling unit from individual codec tokens to codec-token patches.

\subsection{Efficient Decoding for Autoregressive Generation}

Efficient decoding techniques aim to reduce the per-token inference cost of autoregressive generation while preserving the token-level prediction structure.
Multi-token prediction (MTP)~\cite{gloeckle2024better, nguyen2025accelerating, wang2025vocalnet} augments autoregressive models with auxiliary heads that predict multiple future tokens from a shared hidden representation.
Speculative decoding (SD)~\cite{chen2023accelerating, leviathan2023fast} uses a smaller draft model to propose candidate tokens and verifies multiple candidates in parallel with the target model, reducing the effective target-model cost per accepted token.
Recent studies have adapted SD techniques to AR-TTS~\cite{li2025fast, lin2025ssd}.
Although these methods accelerate token-level decoding, they still perform prediction at individual codec-token timesteps, and the KV cache continues to grow with the generated token sequence.
In contrast, TLDR shifts the modeling unit from individual codec tokens to codec-token patches, reducing both the number of global decoding steps and the KV cache maintained by the global sequence model.

\subsection{Patch-level Sequence Modeling}

Patch-level sequence modeling reduces the effective length of long discrete sequences by aggregating low-level tokens into higher-level units.
In the text domain, MEGABYTE~\cite{yu2023megabyte} partitions byte sequences into patches and combines a global model with a local model to autoregressively model long sequences.
BLT~\cite{pagnoni2025blt} further segments bytes into entropy-based dynamic patches and treats patches as the primary units of computation, improving efficiency and robustness in byte-level language modeling.
Hierarchical Transformer architectures, such as Hourglass~\cite{nawrot2022hierarchical} and H-Net~\cite{hwang2025hnet}, also shorten causal sequences through hierarchical or dynamically chunked representations.

These approaches motivate patch-level modeling as a way to reduce sequential computation, but they mainly target text or byte-level language modeling.
For speech generation, DiTAR~\cite{jia2025ditar} introduced a patch-based autoregressive framework that combines a causal language model with a diffusion Transformer over continuous speech representations.
DiTAR therefore differs from TLDR in both representation and integration target: it constructs and trains a new continuous-latent generative pipeline from scratch, whereas TLDR retrofits an existing discrete codec-based AR-TTS system without replacing the pretrained AR backbone or the surrounding TTS pipeline.

%% file: sections/03_method.tex
\section{Method}
\label{sec:method}

\subsection{TLDR Overview}

Let \(x=(x_1,\ldots,x_T)\) denote the target speech token sequence, including the end-of-sequence (EOS) token.
Given a patch size \(k\), we divide the target sequence into \(N=\lceil T/k\rceil\) contiguous patches:
\[
X_i = (x_{(i-1)k+1}, \ldots, x_{\min(ik,T)}),
\qquad i=1,\ldots,N.
\]
If the final patch contains fewer than \(k\) tokens, it is padded for fixed-length patch processing, and padded positions are excluded from the training loss.

\textsc{TLDR} shifts autoregressive modeling from individual speech tokens to the patch sequence \((X_1,\ldots,X_N)\).
The framework consists of three components: (i) a token-to-patch compressor, (ii) a pretrained AR-TTS backbone used as a patch-level global Transformer, and (iii) a speaker-conditioned patch-to-token extractor.
In the zero-shot setting, reference speech tokens are also compressed into prompt patch representations and prepended to the generated patch sequence as acoustic conditioning.

\subsection{Token-to-Patch Compressor}

The compressor maps each patch \(X_i\) to a latent representation \(p_i\), yielding a compressed sequence \((p_1,\ldots,p_N)\).
For each patch, we initialize \(p_i\) by mean-pooling the embeddings of its constituent tokens and applying RMSNorm.
The compressor then refines the patch representations with cross-attention that aggregates local token-level acoustic information.
With fixed-length patches, the compressor aggregates token-level information within each patch while preserving patch boundaries.
Details of the compressor architecture are provided in Appendix~\ref{app:compressor}.

\subsection{Patch-Level Transformer}

The patch-level Transformer uses a pretrained codec-based AR-TTS backbone to model the compressed patch sequence.
It processes one latent representation per patch, reducing the global sequence length from \(T\) speech tokens to approximately \(T/k\) patch positions.
We freeze the pretrained backbone and train LoRA~\cite{hu2022lora} adapters to adapt it to patch-level representations.

For causal patch-level generation, let \(g_i\) denote the global context used to generate the next patch \(X_{i+1}\).
For \(i=0,\ldots,N-1\), \(g_i\) is computed from the conditioning prefix and the previous patch representations \(\{p_1,\ldots,p_i\}\), where \(g_0\) is computed from the prefix alone.
During inference, the global KV cache is updated at the patch rate rather than the speech-token rate.
Details are provided in Appendix~\ref{app:patch_transformer}.

\subsection{Speaker-Conditioned Patch-to-Token Extractor}

The patch-to-token Extractor autoregressively predicts the discrete speech tokens inside each target patch, conditioned on the global patch context \(g_i\) and a speaker embedding \(s\) extracted from the reference speech.
Because patch-level compression may weaken fine-grained speaker characteristics such as timbre and speaking style, we condition the Extractor on \(s\) to preserve speaker identity throughout generation.

The Extractor fuses \(g_i\) and \(s\) into a speaker-conditioned patch context \(c_i\), which conditions its cross-attention layers.
It factorizes the token distribution within each patch as
\[
p(X_{i+1} \mid g_i, s)
=
\prod_{j=1}^{\ell_{i+1}}
p(x_{i+1,j} \mid x_{i+1,<j}, c_i),
\]
where \(\ell_{i+1}\le k\) denotes the number of valid tokens in \(X_{i+1}\), and \(x_{i+1,<j}\) denotes the previously generated tokens within the same target patch.
Thus, the Extractor performs speaker-conditioned autoregressive generation locally within each patch.

We train all trainable modules with teacher-forced token-level cross-entropy over valid target speech tokens.
Details are provided in Appendix~\ref{app:patch_to_token}.

\subsection{Inference}

During inference, \textsc{TLDR} generates speech patch by patch while conditioning on the reference speech prompt.
We compress the prompt speech tokens into patch representations before feeding them to the patch-level Transformer.
This \textit{prompt patchification} aligns the temporal granularity of the prompt with that of the generated patch sequence.
Without this alignment, the patch-level backbone receives the prompt at a mismatched acoustic rate, effectively stretching the prompt by a factor of \(k\) and degrading continuation quality.
The resulting prompt patch representations are used as part of the initial context together with the non-speech prefix and text tokens.

The model then alternates between patch-level context modeling and token-level patch generation until it produces an EOS token or reaches the maximum number of patches.
After decoding, all generated patches are concatenated into a single speech-token sequence and passed to the vocoder for waveform reconstruction.

%% file: sections/04_experiment.tex
\section{Experiments}\label{sec:experiment}
We evaluate \textsc{TLDR} along two axes.
First, we examine whether shifting global causal modeling from token positions to patch positions preserves the intelligibility and speaker similarity of the underlying backbone.
Second, we analyze the quality--latency--memory trade-off induced by varying the patch size \(k\).
We use CosyVoice3~\cite{du2025cosyvoice3} as the reference backbone: because \textsc{TLDR} shares its tokenizer, text frontend, and vocoder, performance differences primarily reflect the proposed patch-level reformulation.

\begin{table}[t]
\centering
\caption{Zero-shot TTS performance on SeedTTS-EN dataset. 
We compare \textsc{TLDR} with AR and NAR baselines using WER, SIM, and RTF.}
\label{tab:main_seed}
\vspace{0.2cm}
\resizebox{1.0\linewidth}{!}{
\setlength\tabcolsep{8pt}
\renewcommand{\arraystretch}{1.1}
\begin{tabular}{clccccc}
\toprule
\multirow{2}{*}{\textbf{Type}} & \multirow{2}{*}{\textbf{Model}} & \multirow{2}{*}{\textbf{Params.}} & \multirow{2}{*}{\textbf{Dataset}} & \multicolumn{3}{c}{\textbf{SeedTTS-EN}} \\
\cmidrule(lr){5-7}
& & & & \textbf{WER} $\downarrow$ & \textbf{SIM} $\uparrow$ & \textbf{RTF} $\downarrow$ \\
\midrule
-- & \textbf{Ground-truth} & - & - & 2.14 & 0.734 & -- \\
\midrule
\multirow{8}{*}{AR}
& Seed-TTS    & --   & --           & 2.25             & \textbf{0.762}    & --    \\
& Spark-TTS~\cite{wang2025sparktts}     & 0.5B & 100k Multi.  & \underline{1.98} & 0.573 & 0.952 \\
& FireRedTTS2~\cite{xie2025fireredtts2} & 1.5B & 248k Multi.  & \textbf{1.95}    & 0.665 & 1.225 \\
& IndexTTS2   & 1.5B & 55k Emilia   & 2.23             & 0.706             & 0.823 \\
& Llasa       & 1B   & 250k Multi.  & 3.22             & 0.572             & --    \\
& CosyVoice   & 0.3B & 170k Multi.  & 4.29             & 0.609             & --    \\
& CosyVoice2  & 0.5B & 170k Multi.  & 2.57             & 0.659             & 0.457 \\
& CosyVoice3  & 0.5B & 1000k Multi. & 2.02             & 0.691             & 0.605 \\
\midrule
\multirow{3}{*}{NAR}
& MaskGCT         & 1.1B & 100k Emilia & 2.62 & \underline{0.714} & -- \\
& E2 TTS (32 NFE) & 0.3B & 100k Emilia & 2.19 & 0.710             & -- \\
& F5-TTS (32 NFE) & 0.3B & 100k Emilia & 2.00 & 0.647             & -- \\
\midrule
\multirow{3}{*}{Ours}
& \textbf{TLDR} ($k{=}4$) & 0.5B + 136.2M & 0.6k LibriTTS & 2.03 & 0.684 & 0.336 \\
& \textbf{TLDR} ($k{=}6$) & 0.5B + 136.2M & 0.6k LibriTTS & 2.10 & 0.686 & \underline{0.278} \\
& \textbf{TLDR} ($k{=}8$) & 0.5B + 136.2M & 0.6k LibriTTS & 2.49 & 0.688 & \textbf{0.248} \\
\bottomrule
\end{tabular}}
\end{table} 

\subsection{Experimental Setup}\label{sec:setup}

\paragraph{Training setup.}
We train \textsc{TLDR} on 585 hours of LibriTTS~\cite{zen2019libritts}, using the CosyVoice3 speech tokenizer (25 tokens/sec, \(V=6{,}561\)) and the Qwen2 text tokenizer~\cite{yang2024qwen2}.
For speaker conditioning, we extract speaker embeddings from the reference speech using WavLM~\cite{chen2022wavlm} with an ECAPA-TDNN head~\cite{desplanques2020ecapa}.
The patch-level Transformer is initialized from the CosyVoice3 AR backbone, kept frozen, and adapted only through LoRA adapters.
The token-to-patch compressor and patch-to-token extractor are trained from scratch.
This results in 136.2M trainable parameters out of 648M total parameters.
We report results for \(k\in\{4,6,8\}\); more details are provided in Tables~\ref{tab:model_config} and~\ref{tab:training_config}.

\paragraph{Evaluation protocol.}
We evaluate zero-shot TTS on SeedTTS-EN ~\cite{anastassiou2024seed} and LibriSpeech-PC   subset B~\cite{panayotov2015librispeech}, containing 1{,}088 and 1{,}127 utterances, respectively.
For objective evaluation, we report word error rate (WER), speaker similarity (SIM), UTMOS~\cite{saeki2022utmos}, and real-time factor (RTF) where applicable.
WER is computed by transcribing generated speech with \texttt{whisper large-v3}~\cite{radford2023robust} and comparing the transcripts against the ground-truth text.
SIM is computed between the prompt and synthesized speech using WavLM-TDNN~\cite{chen2022wavlm}.
To check that the SIM results are not specific to this evaluator, we additionally report speaker similarity with an independent WeSpeaker model in Appendix~\ref{app:independent_sim}.

\paragraph{Subjective evaluation.}
We additionally conduct a crowdsourced subjective evaluation with 25 English speakers.
SMOS (similarity mean opinion score) measures perceived speaker similarity to the prompt on a 1--5 scale with 0.5-point increments.
For naturalness, raters compare \textsc{TLDR} against the CosyVoice3 baseline using both CMOS (comparative mean opinion score) on a $-3$ to $+3$ scale and an A/B preference test; positive CMOS scores and A/B preferences indicate preference for \textsc{TLDR}. 

\paragraph{Inference protocol.}
We measure RTF on a single A100 80GB GPU with batch size 1, using a fixed 100-token generation corresponding to approximately 4 seconds of audio.
Inference memory is measured with batch size 64 under fp16 SDPA across output durations of 4--20 seconds.
All systems use the same sampling configuration: temperature \(0.7\), top-\(k\) \(25\), and top-\(p\) \(0.7\).

\paragraph{Baseline scores.}
Unless otherwise noted, the baseline results in Tables~\ref{tab:main_seed} and~\ref{tab:main_libri} are taken from the official CosyVoice3 repository.
We evaluate F5-TTS and E2 TTS under matched decoding settings, and take the DiTAR result in Table~\ref{tab:main_libri} from the original paper~\cite{jia2025ditar}.

\begin{table}[t]
\centering
\caption{Zero-shot TTS performance on \emph{LibriSpeech-PC}  .
We compare \textsc{TLDR} with AR and NAR baselines using WER, SIM, and UTMOS.}
\vspace{0.2cm}
\label{tab:main_libri} 
\resizebox{1.0\linewidth}{!}{
\setlength\tabcolsep{8pt}
\renewcommand{\arraystretch}{1.1}
\begin{tabular}{@{}clcccccc@{}}
\toprule
\multirow{2}{*}{\textbf{Type}} 
& \multirow{2}{*}{\textbf{Model}} 
& \multirow{2}{*}{\textbf{Params.}} 
& \multirow{2}{*}{\textbf{Dataset}}
& \multicolumn{3}{c}{\textbf{LibriSpeech-PC  }} \\
\cmidrule(lr){5-7}
& & & & \textbf{WER} $\downarrow$ & \textbf{SIM} $\uparrow$ & \textbf{UTMOS} $\uparrow$ \\
\midrule
-- & \textbf{Ground-truth} & -- & -- & 2.23 & 0.69 & 4.10 \\
\midrule
\multirow{5}{*}{AR}
& CosyVoice  & 0.3B & 170k Multi.  & 3.59             & 0.66              & 4.14              \\
& CosyVoice2 & 0.5B & 170k Multi.  & \underline{2.05} & 0.657             & \textbf{4.38}     \\
& FireRedTTS & 0.6B & 248k Multi.  & 2.69             & 0.47              & --                \\
& DiTAR      & 0.6B & 100k Emilia  & 2.39             & 0.67              & 4.22              \\
& CosyVoice3 & 0.5B & 1000k Multi. & \textbf{1.95}    & \textbf{0.718}    & \underline{4.28}  \\
\midrule
\multirow{3}{*}{NAR}
& MaskGCT (50 NFE) & 1.1B & 100k Emilia & 2.72 & 0.69 & 3.90 \\
& E2 TTS (32 NFE)  & 0.3B & 100k Emilia & 2.95 & 0.69 & 3.56 \\
& F5-TTS (32 NFE)  & 0.3B & 100k Emilia & 2.42 & 0.66 & 3.88 \\
\midrule
\multirow{3}{*}{Ours}
& \textsc{TLDR} ($k{=}4$) & 0.5B + 136.2M & 0.6k LibriTTS & 2.15 & \underline{0.710} & 4.24 \\
& \textsc{TLDR} ($k{=}6$) & 0.5B + 136.2M & 0.6k LibriTTS & 2.20 & 0.708             & 4.23 \\
& \textsc{TLDR} ($k{=}8$) & 0.5B + 136.2M & 0.6k LibriTTS & 2.53 & 0.709             & 4.23 \\
\bottomrule
\end{tabular}}
\end{table}   

\subsection{Experimental Results}\label{sec:result}
\paragraph{Zero-shot TTS Quality.}\label{subsec:quality}

Tables~\ref{tab:main_seed} and~\ref{tab:main_libri} summarize the main zero-shot TTS results.
Overall, \textsc{TLDR} preserves much of the CosyVoice3 backbone's quality while using a substantially smaller adaptation set.
With $k{=}4$, \textsc{TLDR} closely matches CosyVoice3 on SeedTTS-EN and remains competitive with strong AR baselines on LibriSpeech-PC.
These results suggest that patch-level decoding can improve efficiency without substantially degrading zero-shot quality.

We also conduct a subjective evaluation on the SeedTTS-EN.
As shown in Table~\ref{tab:subjective}, \textsc{TLDR} ($k{=}4$) obtains comparable speaker-similarity ratings and slightly higher naturalness preference than CosyVoice3.
We further validate the role of speaker conditioning in Section~\ref{subsec:abl_speaker}.

\begin{figure}[!t]
\centering
\includegraphics[width=0.95\textwidth]{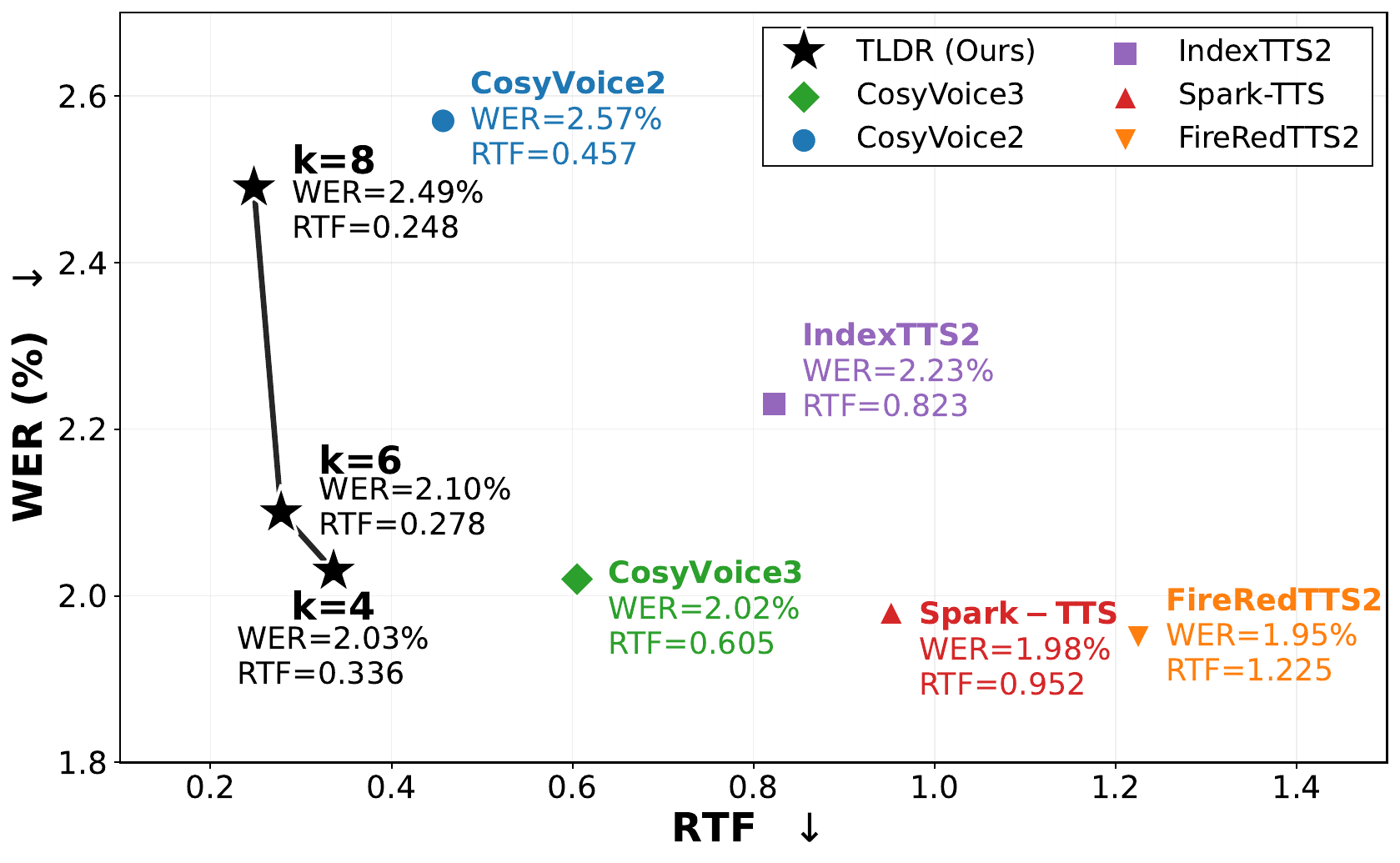}
\caption{
WER--RTF trade-off of \textsc{TLDR} with different patch sizes, compared with existing speech generation models.
}
\label{fig:tradeoff}
\end{figure}

\paragraph{WER--RTF Trade-off.}\label{subsec:tradeoff}
Figure~\ref{fig:tradeoff} compares the WER--RTF trade-off of \textsc{TLDR} under different patch sizes against existing speech generation models.
Larger patches reduce the number of global autoregressive decoding steps, lowering RTF.
However, they also require the local decoder to predict more speech tokens within each patch, which can hurt WER.
Since speaker similarity and perceptual quality remain relatively stable across patch sizes, increasing $k$ mainly trades content accuracy for faster inference.

\paragraph{Memory Efficiency.}\label{subsec:memory}

\begin{figure}[t]
\centering
\includegraphics[width=\textwidth]{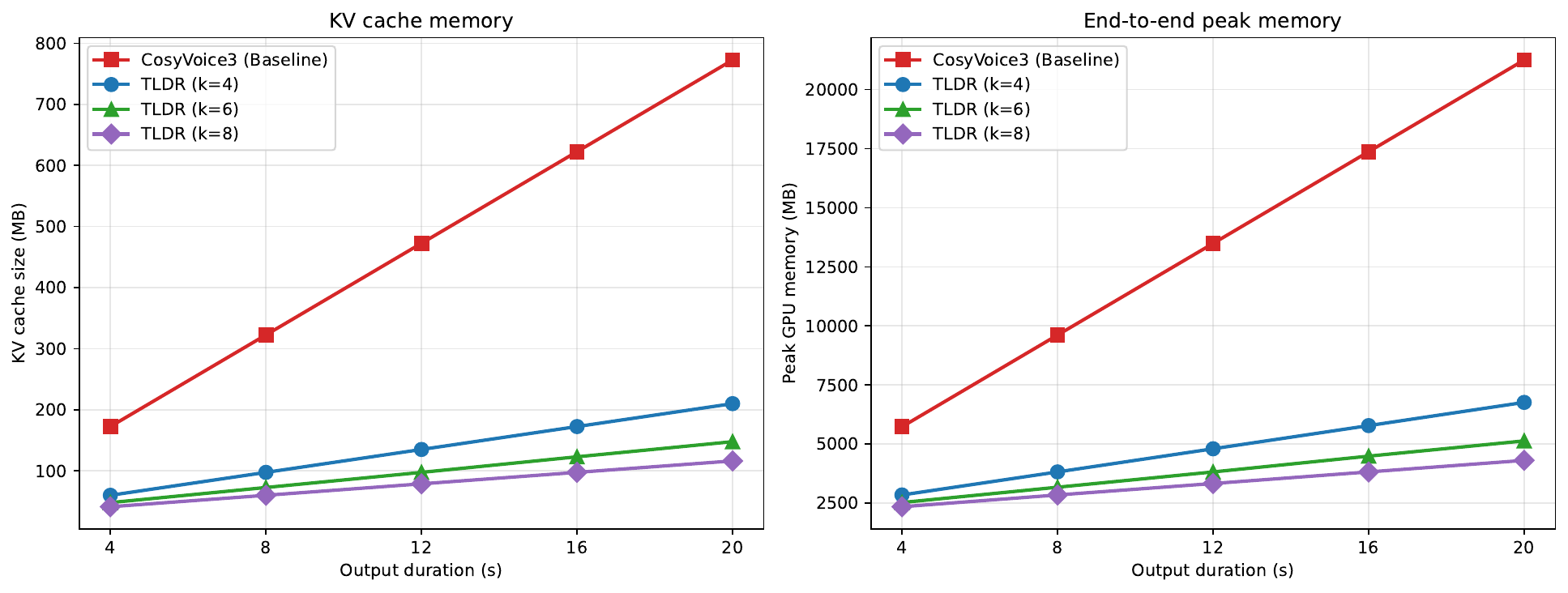}
\caption{
Inference memory comparison between \textsc{TLDR} and the CosyVoice3 backbone across output durations at batch size 64.
(Left) global KV cache size; (Right) end-to-end peak GPU memory.
}
\label{fig:memory}
\end{figure}

Figure~\ref{fig:memory} shows that the efficiency benefits of patch-level decoding extend beyond decoding latency to memory.
Because the global Transformer caches one KV state per patch instead of per speech token, the global KV cache grows as $T/k$ rather than $T$, and the gap relative to CosyVoice3 widens with output duration (see Table~\ref{tab:mtp_ablation}).
End-to-end peak GPU memory at batch~64 also decreases, even after accounting for the added compressor and extractor parameters. 
This is because the dominant memory cost, the global KV cache, grows at the patch rate rather than the token rate.
This is where \textsc{TLDR}'s structural reformulation matters most: in batched serving, KV-cache occupancy rather than parameter count is typically the binding constraint~\cite{kwon2023efficient}, so per-patch caching directly translates into higher achievable batch size and throughput.

\begin{table}[t]
\centering
\caption{Subjective evaluation on SeedTTS-EN using SMOS, CMOS, and A/B preference.}
\label{tab:subjective}
\vspace{0.2cm}
\renewcommand{\arraystretch}{1.1}
\begin{tabular}{lccc}
\toprule
\textbf{System} & \textbf{SMOS} ($\uparrow$) & \textbf{CMOS} ($\uparrow$) & \textbf{A/B Preference (\%)}\\
\midrule 
CosyVoice3              &  3.739 & 0.00 & 46.1\% \\
\textbf{TLDR} ($k{=}4$) & 3.953  &  0.19 & 53.9\%\\
\bottomrule
\end{tabular}
\end{table}  

%% file: sections/05_analysis.tex
\section{Analysis}
\label{sec:analysis} 
We analyze the main design choices of \textsc{TLDR} on SeedTTS-EN.
Unless otherwise noted, all variants use $k{=}4$ and share the same tokenizer, text frontend, vocoder, and training data, so that differences reflect the ablated component. 

We study LoRA adaptation of the frozen backbone, speaker conditioning in the patch-to-token extractor, and within-patch parallelization. 
The compressor cross-attention ablation is provided in Appendix Table~\ref{tab:compressor_ablation}.

\subsection{LoRA Adaptation of the Frozen Backbone}
\label{subsec:abl_lora}

We test whether the frozen CosyVoice3 backbone can be reused directly with patch-level inputs.
Although the backbone already contains strong autoregressive speech modeling capability, it is pretrained on token-level sequences rather than compressed patch representations.
We compare the full model with a variant that removes LoRA and trains only the token-to-patch compressor and patch-to-token extractor.

Table~\ref{tab:lora_ablation} shows that removing LoRA substantially degrades WER while leaving SIM nearly unchanged.
This suggests that LoRA mainly adapts the frozen backbone to the patch-level input space needed for content prediction.
Without this adaptation path, the compressor and extractor alone do not sufficiently bridge the mismatch between token-level pretraining and patch-level inputs.

\begin{figure}[t!] 
\centering 
\includegraphics[width=0.9\linewidth]{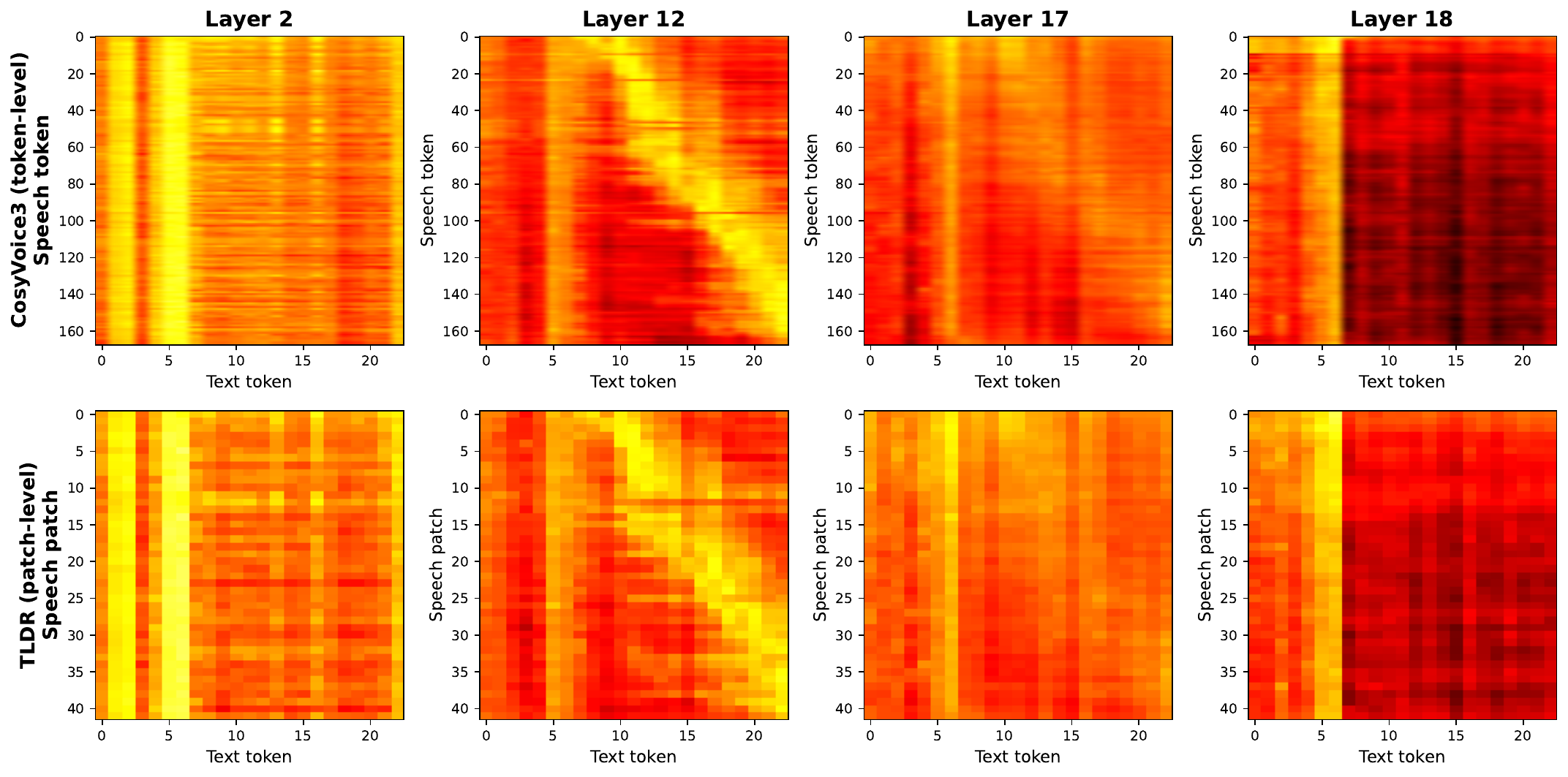} 
\caption{
Text--speech cross-attention maps for CosyVoice3 and \textsc{TLDR} ($k{=}4$).
Despite the compressed speech axis, \textsc{TLDR} preserves the left-to-right text--speech alignment observed in the token-level backbone.
}\label{fig:compare_layer} 
\end{figure}

\subsection{Speaker Conditioning}
\label{subsec:abl_speaker}

We examine the role of reference-speaker conditioning in the patch-to-token
extractor.
In the ablated variant, we remove the speaker embedding $s$ and condition the local generator only on the global patch context $g_i$.

Table~\ref{tab:speaker_conditioning_ablation} shows that removing speaker conditioning improves WER but substantially reduces SIM.
This trade-off suggests that the global patch context is sufficient for content prediction, while explicit speaker conditioning plays a central role in preserving the reference speaker under patch-level generation.
Since zero-shot TTS requires both intelligibility and speaker similarity, we keep speaker conditioning in the main model.

  \begin{table}[t]
  \begin{minipage}[b]{0.48\textwidth}                                                                                                                       
  \centering      
  \setlength\tabcolsep{4pt}
  \caption{Effect of LoRA adaptation on SeedTTS-EN.}  
  \vspace{0.2cm}
  \label{tab:lora_ablation}
  \begin{tabular*}{\linewidth}{@{\extracolsep{\fill}}lcc}                                                                                                   
  \toprule                                                                                                                                                  
  Backbone adaptation & WER (\%) $\downarrow$ & SIM $\uparrow$ \\
  \midrule                                                                                                                                                  
  No LoRA          & 7.66 & 0.687 \\
  $+$ LoRA (Ours)  & 2.03 & 0.684 \\                                                                                                                        
  \bottomrule
  \end{tabular*}                                                                                                                                            
  \end{minipage}  
  \hfill                                                                                                                                                    
  \begin{minipage}[b]{0.48\textwidth}
  \centering                                                                                                                                                
  \setlength\tabcolsep{4pt}
  \caption{Effect of speaker conditioning on SeedTTS-EN.}
  \vspace{0.2cm}
  \label{tab:speaker_conditioning_ablation}                                                                                                                 
  \begin{tabular*}{\linewidth}{@{\extracolsep{\fill}}lcc}
  \toprule                                                                                                                                                  
  Model variant & WER (\%) $\downarrow$ & SIM $\uparrow$ \\
  \midrule                                                                                                                                                  
  w/o speaker cond. & 1.77 & 0.580 \\
  w/  speaker cond. & 2.03 & 0.684 \\                                                                                                                       
  \bottomrule
  \end{tabular*}                                                                                                                                            
  \end{minipage}  
  \end{table}

\subsection{Patch-Level Reformulation vs.\ Within-Patch Parallelization}
\label{subsec:abl_mtp}

We compare \textsc{TLDR} with variants that parallelize token prediction within
each patch.
This comparison distinguishes the effect of shortening the global autoregressive
sequence from the effect of parallelizing the local patch-to-token decoding.
The multi-token prediction (MTP) variant predicts multiple future tokens inside
each patch using additional heads, while the NAR extractor predicts all tokens in
a patch in parallel.

Table~\ref{tab:mtp_ablation} shows that \textsc{TLDR} reduces the global KV cache
to \(0.26\times\) that of CosyVoice3 at 12 seconds, while maintaining comparable
WER and SIM.
The MTP and non-autoregressive extractor variants use the same patch-level global
sequence as \textsc{TLDR}, and therefore have the same global KV-cache ratio.
However, both variants degrade WER and SIM, indicating that parallelizing
within-patch prediction hurts intelligibility and speaker similarity.

These results suggest that the main memory reduction comes from running the
global backbone at the patch rate, while autoregressive decoding inside each
patch remains important for preserving intelligibility and speaker similarity.

\begin{table}[t]
\centering
\caption{Comparison of patch-level reformulation and within-patch parallelization on SeedTTS-EN.}
\vspace{0.2cm}
\label{tab:mtp_ablation}
\resizebox{1.0\linewidth}{!}{
\begin{tabular}{lccc}
\toprule
Variant & WER (\%) $\downarrow$ & SIM $\uparrow$ & Global KV cache ratio @12\,s $\downarrow$ \\
\midrule
CosyVoice3                                            & 2.02 & 0.691 & 1.00$\times$ \\
\textsc{TLDR} ($k{=}4$)                               & 2.03 & 0.684 & 0.26$\times$ \\
\textsc{TLDR} ($k{=}4$) $+$ MTP (3 heads)             & 4.85 & 0.631 & 0.26$\times$ \\
\textsc{TLDR} ($k{=}4$), NAR Patch-to-Token Extractor & 3.79 & 0.573 & 0.26$\times$ \\
\bottomrule
\end{tabular}}
\end{table}
 
\subsection{Text--Speech Alignment under Patch-Level Decoding}
\label{subsec:alignment_analysis} 
Because speech evolves continuously over time, adjacent codec tokens often encode locally correlated acoustic patterns.
This locality suggests that grouping neighboring tokens into patches can preserve the coarse text--speech alignment of the pretrained backbone, while representing speech at a lower temporal resolution.
We test this hypothesis by visualizing text--speech cross-attention maps for \textsc{TLDR} and CosyVoice3 on the same utterance.

Figure~\ref{fig:compare_layer} compares the cross-attention maps of CosyVoice3 and \textsc{TLDR}.
In later layers, both models show a monotonic text--speech alignment: earlier speech positions attend to earlier text tokens, while later speech positions attend to later text tokens.
The difference lies in the speech-axis resolution.
In CosyVoice3, each row corresponds to an individual speech token, whereas in \textsc{TLDR}, each row corresponds to a speech patch.

This qualitative comparison supports our hypothesis that patch-level decoding preserves the coarse left-to-right alignment learned by the pretrained backbone.
Rather than collapsing the alignment, \textsc{TLDR} represents a similar text--speech alignment structure at a coarser temporal granularity.

%% file: sections/06_conclusion.tex
\section{Conclusion}\label{sec:conclusion}

In this work, we proposed TLDR, a patch-based autoregressive framework that reduces the inference cost of pretrained codec-based AR-TTS systems.
TLDR decouples acoustic tokenization from the temporal resolution of global causal modeling: speech remains represented as discrete codec tokens, while the frozen backbone models a shorter sequence of patch-level representations.
By retaining the existing tokenizer, text frontend, vocoder, and pretrained backbone, TLDR retrofits an existing AR-TTS pipeline rather than training a new system from scratch.
Patch size exposes an explicit quality--latency--memory control variable.
At $k=4$, TLDR reduces global-backbone KV-cache memory by approximately 75\% and lowers RTF from 0.605 to 0.336, corresponding to a 1.8$\times$ speedup, with minimal absolute WER increase and a SIM decrease relative to the baseline.

%% file: sections/Y_appendix.tex
\section{Implementation Details}

\begin{table}[h]
\centering
\caption{Model configuration of \textsc{TLDR}.}
\label{tab:model_config}
\small
\begin{tabular}{llc}
\toprule
\textbf{Component} & \textbf{Hyperparameter} & \textbf{Value} \\
\midrule
\multirow{4}{*}{Token-to-Patch Compressor}
  & Layers & 1 \\
  & Hidden dim / Heads & 896 / 8 \\
  & FFN dim & 3584 \\
  & Cross-attn heads & 4 \\
\midrule
\multirow{3}{*}{Patch-Level Transformer}
  & Backbone & Qwen2-0.5B \\
  & LoRA rank / alpha & 64 / 64 \\
  & Hidden dim & 896 \\
\midrule
\multirow{5}{*}{Patch-to-Token Extractor}
  & Layers & 4 \\
  & Hidden dim / Heads & 896 / 8 \\
  & FFN dim & 3584 \\
  & Cross-attn heads & 4 \\
  & Slots per layer ($m$) & 4 \\
\midrule
\multirow{2}{*}{Patching}
  & Patch size ($k$) & 4 \\
  & Local window ($w$) & 16 \\
\bottomrule
\end{tabular}
\end{table}

\begin{table}[h]
\centering
\caption{Training configuration of \textsc{TLDR}.}
\label{tab:training_config}
\small
\begin{tabular}{ll}
\toprule
\textbf{Hyperparameter} & \textbf{Value} \\
\midrule
Optimizer                                & AdamW \\
Learning rate (Compressor / Extractor)   & $1\mathrm{e}{-4}$ \\
Learning rate (LoRA on Patch-Level Transformer) & $5\mathrm{e}{-5}$ \\
Weight decay                             & $0.01$ \\
LR schedule                              & Linear warmup $\to$ cosine decay \\
Warmup steps                             & $5{,}000$ \\
Final LR ratio                           & $0.1$ \\
Per-device batch size                    & $16$ \\
Gradient accumulation                    & $4$ \\
Effective batch size                     & $64$ \\
Max gradient norm                        & $1.0$ \\
Max training steps                       & $500{,}000$ \\
Maximum sequence length                  & $512$ \\
Dropout                                  & $0.1$ \\
Random seed                              & $42$ \\
Hardware                                 & 1$\times$ NVIDIA A100 80\,GB \\
Training time                            & $\sim$20 hours \\
\bottomrule
\end{tabular}
\end{table}

\subsection{Details of the Token-to-Patch Compressor}
\label{app:compressor}

After mean-pooling and RMSNorm initialization, the token-to-patch compressor refines patch representations using a block that interleaves a local Transformer layer with patch-to-token cross-attention.
The local Transformer layer contextualizes token hidden states with windowed causal self-attention (RoPE positional encoding) and a SwiGLU feed-forward network.
The patch representations are then updated through masked patch-to-token cross-attention, where each patch representation serves as a query and attends only to token hidden states assigned to the same patch.

The cross-attention mask ensures that patch \(i\) only aggregates information from tokens in \(X_i\), preventing information from being mixed across patch boundaries.

\subsection{Details of the Patch-Level Transformer}
\label{app:patch_transformer}

We initialize the Patch-Level Transformer from a CosyVoice3-style codec-based AR-TTS checkpoint built on Qwen2-0.5B.
LoRA adapters are inserted into the frozen backbone as the trainable parameters in the global path; rank and alpha values are reported in Table~\ref{tab:model_config}.

The global input sequence is formed by concatenating the non-speech prefix---consisting of SOS, text, and task tokens---with the compressed patch representations.
During training, we use a shifted patch sequence so that the context $g_i$ used to predict target patch $X_{i+1}$ is produced from the prefix and previous patch representations $p_1,\ldots,p_i$, without exposing the target patch representation $p_{i+1}$.
This shift is necessary because $p_{i+1}$ is computed from the tokens in $X_{i+1}$ during teacher-forced training; using it to predict $X_{i+1}$ would introduce target leakage.

\subsection{Details of the Speaker-Conditioned Patch-to-Token Extractor}
\label{app:patch_to_token}

Given a speaker embedding \(s\) extracted from the reference speech, we project it to \(v_s = W_s s\).
We then concatenate \(v_s\) with each global patch context \(g_i\) and obtain a speaker-conditioned patch context
\[
\tilde g_i = W_c [g_i; v_s].
\]
Each decoder layer then projects \(\tilde g_i\) into \(m\) cross-attention slots:
\[
C_i^{(\ell)}
=
\operatorname{reshape}
\left(
W_{\mathrm{slot}}^{(\ell)} \tilde g_i
\right)
\in \mathbb{R}^{m \times d},
\]
where \(d\) is the hidden dimension of the patch-to-token extractor and \(m\) is the number of slots per layer (Table~\ref{tab:model_config}).

At each patch-to-token extractor layer, token representations first attend to the speaker-conditioned slots through cross-attention, followed by causal self-attention within the current patch.
Training uses teacher forcing inside each target patch with a token-level cross-entropy loss over valid target speech-token positions, excluding prompt and padding tokens.

\section{Cross-Attention in the Token-to-Patch Compressor}
\label{subsec:abl_compressor}

We test whether the token-to-patch compressor needs a learned cross-attention module or whether a simple average of token embeddings is sufficient.
We compare the full \textsc{TLDR} model with a variant that replaces the compressor with mean pooling followed by RMSNorm.
All other components, including the LoRA adapters and the patch-to-token extractor, are kept unchanged.

Table~\ref{tab:compressor_ablation} shows that replacing the cross-attention compressor with mean pooling increases WER from \(2.03\)\% to \(2.66\)\%, while SIM remains unchanged.
This suggests that the compressor mainly affects content modeling rather than speaker preservation.
Speaker similarity is not affected because the speaker embedding is injected in the patch-to-token extractor, outside the global patch-level path.
In contrast, content accuracy depends on the patch representation passed to the global backbone.
Mean pooling gives each token in a patch the same weight, whereas cross-attention learns which token-level features should be passed to the patch representation.
We therefore use the cross-attention compressor in the main model.

\begin{table}[t]
\centering
\caption{Effect of the Token-to-Patch Compressor on SeedTTS-EN test-clean.
We compare mean pooling with the cross-attention compressor used in \textsc{TLDR}.}
\vspace{0.2cm}
\label{tab:compressor_ablation}
\begin{tabular}{lcc}
\toprule
Patch representation & WER (\%) $\downarrow$ & SIM $\uparrow$ \\
\midrule
Mean pool $+$ RMSNorm only                  & 2.66 & 0.6842 \\
Cross-attention Compressor (\textsc{TLDR})  & 2.03 & 0.6842 \\
\bottomrule
\end{tabular}
\end{table}

\section{Speaker Similarity with an Independent Evaluator}
\label{app:independent_sim}

Our main speaker-similarity evaluation uses a WavLM encoder with an ECAPA-TDNN
head~\cite{chen2022wavlm, desplanques2020ecapa}, which is the same speaker
embedding model used for speaker conditioning.
To assess whether the reported SIM scores depend on this embedding space, we
additionally evaluate speaker similarity using an independent speaker
verification model based on WeSpeaker~\cite{wang2023wespeaker}.

Table~\ref{tab:wespeaker_sim} shows the same trend:
\textsc{TLDR} achieves a speaker similarity score comparable to CosyVoice3 under
the independent evaluator.
This suggests that the speaker-similarity comparison is not specific to the
WavLM-ECAPA embedding space used for conditioning.

\begin{table}[t]
\centering
\caption{Speaker similarity evaluated with an independent WeSpeaker SIM model
on SeedTTS-EN. The results are reported as mean $\pm$ standard deviation.}
\label{tab:wespeaker_sim}
\vspace{0.2cm}
\begin{tabular}{lc}
\toprule
Model & WeSpeaker SIM $\uparrow$ \\
\midrule
CosyVoice3 & $0.7699 \pm 0.073$ \\
\textsc{TLDR} ($k{=}4$) & $0.7619 \pm 0.067$ \\
\bottomrule
\end{tabular}
\end{table}

\section*{Limitations and Future Work} 
\textsc{TLDR} is studied in a controlled single-backbone setting, leaving several extensions for future work.
We evaluate only one AR-TTS backbone, CosyVoice3, adapted on 585 hours of LibriTTS. Future work will test other public AR-TTS backbones, multilingual and larger-scale data, and long-form generation.
The current model uses a fixed patch size $k$, despite the uneven information density of speech. Adaptive patching could assign larger patches to redundant regions and smaller patches to dense transitions, improving the speed--quality trade-off.
Our subjective test uses 25 English listeners, sufficient for backbone comparison but limited in scale and language coverage. Future work will expand both.                                                                                                
\section*{Broader Impact}
\textsc{TLDR} reduces the inference cost of codec-based AR-TTS, lowering energy consumption and hardware requirements for high-quality speech synthesis. This can improve accessibility on commodity or resource-constrained hardware. At the same time, \textsc{TLDR}-accelerated systems inherit the risks of TTS systems, including voice forgery. Standard audio-watermarking and deepfake-detection methods remain applicable, and we encourage their use in deployment.